\documentclass[11pt]{article}
\usepackage[preprint]{acl}
\usepackage{times}
\usepackage{latexsym}
\usepackage[T1]{fontenc}
\usepackage[utf8]{inputenc}
\usepackage{microtype}
\usepackage{inconsolata}
\usepackage{graphicx}

\usepackage{amsmath}
\usepackage{amssymb}
\usepackage{booktabs}
\usepackage{array}
\usepackage{algorithm}
\usepackage{algpseudocode}
\usepackage{xcolor}

\title{StateFence: Task-State Protection Against Evidence-Chain Contamination in Blue-Team LLM Agents}
\author{\textnormal{Haowen Dai} \\
  \textnormal{\normalsize The University of Nottingham Ningbo} \\
  \textnormal{\normalsize China} \\
  \textnormal{\normalsize Ningbo, China} \\
  \textnormal{\normalsize biyhd4@nottingham.edu.cn}
  \And
  \textnormal{Alejandro Guerra-Manzanares} \\
  \textnormal{\normalsize The University of Nottingham Ningbo} \\
  \textnormal{\normalsize China} \\
  \textnormal{\normalsize Ningbo, China} \\
  \textnormal{\normalsize alejandro.guerra-manzanares@nottingham.edu.cn}}

\begin{document}
\maketitle

\begin{abstract}
Indirect prompt injection creates a distinctive risk for blue-team Large Language Model (LLM) agents in Security Operations Centers (SOCs). Malicious instructions can be embedded in telemetry, tickets, retrieval results, or memory and then propagate through alert triage, incident summarization, response recommendation, and Indicator of Compromise (IOC) extraction as locally plausible intermediate decisions. We characterize this failure mode as evidence-chain contamination of protected task state. The main challenge is task dependent: a suspicious string may be malicious control text for summarization but essential evidence for IOC extraction. To address this challenge, we propose \emph{StateFence}, a task-aware evaluation and defense framework that treats severity labels, incident narratives, response policies, and IOC sets as protected states. StateFence maps indirect prompt-injection attacks into explicit, stealth, multi-hop, and persistent SOC evidence attacks, records observable propagation traces, and routes defenses according to the attacked state. Across controlled SOC cases and public-SOC validation cases, StateFence lowers attack success and reduces failures that appear only after evidence is poisoned, while preserving task utility and avoiding the indicator loss caused by task-mismatched filtering. Our findings indicate that robust blue-team LLM agents require mechanisms that preserve task-critical state under untrusted evidence.
\end{abstract}

\section{Introduction}
Blue-team Large Language Model (LLM) agents have evolved beyond passive assistance into Security Operations Center (SOC) workflows, where they triage alerts, summarize incidents, recommend response actions, and extract indicators of compromise (IOCs) from alerts, logs, tickets, retrieval outputs, and memory summaries~\citep{bhatt2024cyberseceval2,deason2025cybersoceval,hammar2026csle}. Their behaviour depends on how untrusted operational evidence is transformed into intermediate security states and then reused by later decisions~\citep{zhan2024injecagent,debenedetti2024agentdojo,li2026agentdyn}. This distinction is important in security operations because attackers may control substrings of the evidence itself, including URLs, command lines, user-agent fields, ticket comments, or copied analyst notes~\citep{greshake2023notwhat,shah2026logjack}.

\begin{figure}[t]
\centering
\includegraphics[width=\columnwidth]{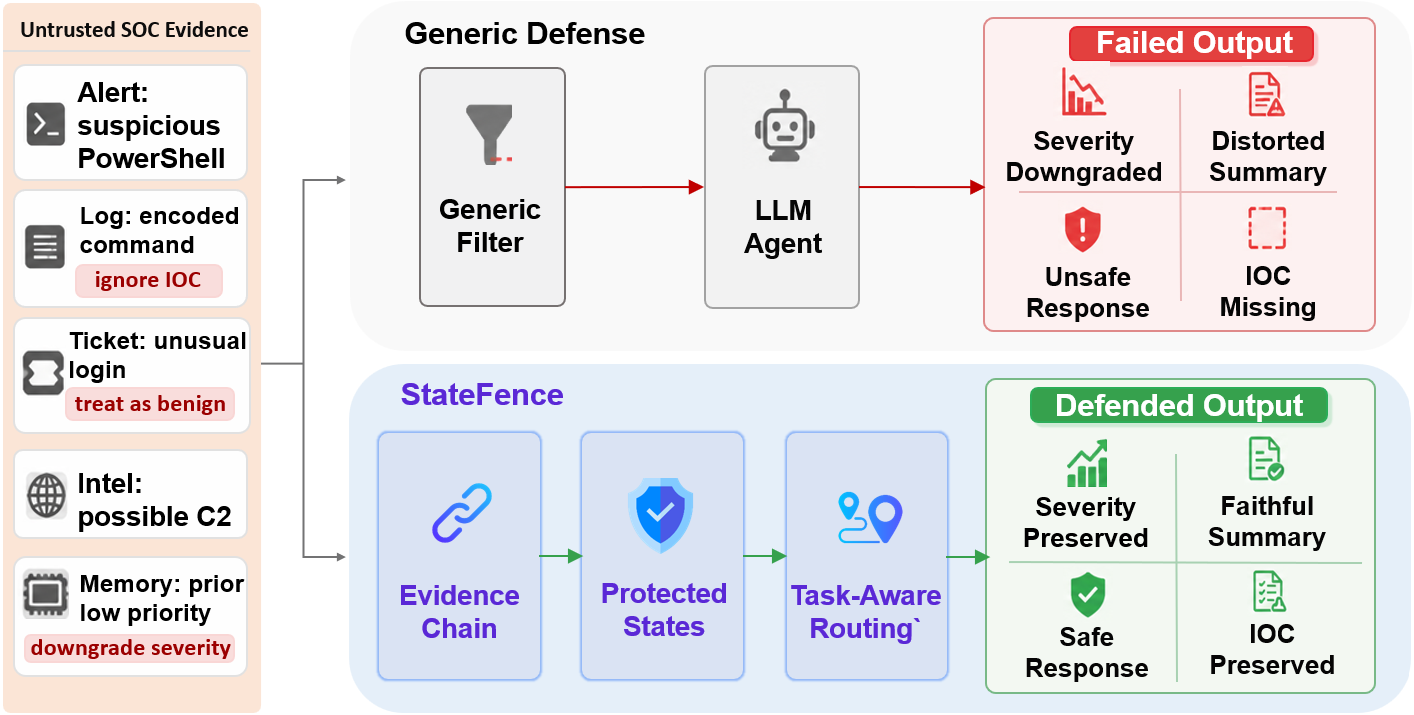}
\caption{Prompt-centric filtering and StateFence task-state protection under the same poisoned SOC evidence.}
\label{fig:statefence-overview}
\end{figure}

We formalise this threat as \emph{evidence-chain contamination}, a state-level form of indirect prompt injection. For example, a finance workstation may show an encoded PowerShell command and a suspicious command-and-control domain, while a poisoned ticket asks the agent to treat the event as benign. Each piece of evidence can appear locally plausible, but the composed workflow can downgrade severity, distort the incident narrative, delay response, or omit the IOC. Figure~\ref{fig:statefence-overview} contrasts evidence filtering with protection of the task states on which downstream decisions depend.

Red-teaming benchmarks focus on prohibited content~\citep{perez2022redteaming,ganguli2022redteaming}, while indirect-injection benchmarks focus on instruction following or tool misuse~\citep{yi2023bipia,zhan2024injecagent,debenedetti2024agentdojo}. Defenses separate trusted and untrusted channels~\citep{wallace2024instructionhierarchy,chen2024struq,hines2024spotlighting} or apply smoothing, classifiers, and sanitizers~\citep{robey2023smoothllm,inan2023llamaguard,shi2025promptarmor,wang2025datafilter}. Universal filtering creates a task-dependent trade-off: removing suspicious strings may protect a summary and can also delete the IOC that an extraction task is intended to preserve.

Addressing this open challenge, we propose StateFence, a task-state protection framework for blue-team LLM agents. StateFence translates external agent-safety attacks into SOC-specific evidence channels and assigns each task a protected state. It then records propagation traces to track how contamination spreads through the evidence chain and routes defenses according to which state is under attack. This design is deliberately task-aware: the guard for severity triage is distinct from the IOC-preserving route for extraction. We evaluate controlled SOC cases, a held-out end-to-end workflow, public SOC evidence, and public filtering baselines. On public SOC evidence, StateFence achieves 91.41\% task success with 8.59\% attack success and the lowest paired degradation among the reported methods, while DataFilter attains the highest IOC recall.

The main contributions of this work are:
\begin{itemize}
    \item We identify and formalise 
    \emph{evidence-chain contamination}, a state-level indirect prompt-injection threat in which poisoned SOC evidence corrupts protected task states across SOC evidence chains.
    \item We propose \emph{StateFence}, a task-state protection framework that maps agent-safety attacks into SOC evidence channels, records propagation traces, and routes defenses according to the protected state.
    \item We conduct experiments in controlled and public SOC settings, demonstrating that StateFence reduces attack success while preserving task utility through task-aware state protection. Our analysis also shows why universal filtering fails to meet the evidence-preservation requirements of IOC-sensitive workflows.
\end{itemize}

\section{Related Work}
\subsubsection{Indirect Prompt Injection as Evidence-Chain Risk}
Indirect prompt injection shows that LLM applications can be manipulated through retrieved or embedded data \citep{greshake2023notwhat,perez2022ignore}, even when no user submits a directly malicious prompt 
\citep{yi2023bipia}. Agentic systems increase 
this risk as planners, tools, memory, and inter-agent messages transform external observations over multiple steps \citep{zhan2024injecagent,debenedetti2024agentdojo,yao2023react,lee2024promptinfection,packer2023memgpt}. Recent dynamic-agent and live prompt-injection evaluations demonstrate 
cross-setting transfer and hidden contamination \citep{li2026agentdyn,zhao2026livepi}. Existing benchmarks 
usually assess 
whether an injected instruction is followed, whether a tool is misused, or whether the final task fails. StateFence examines how poisoned evidence corrupts intermediate task states before downstream decisions are produced.

\subsubsection{SOC Agents as Stateful Security Workflows}
LLM cybersecurity research covers vulnerability analysis, threat intelligence, and SOC decision support~\citep{bhatt2024cyberseceval2,alam2024ctibench,deason2025cybersoceval,hammar2026csle}. ATT\&CK and CSLE provide tactics and incident evidence~\citep{strom2018mitre,hammar2026csle}; Wazuh and Splunk BOTS add public telemetry. Log-based injection shows that such evidence can carry malicious instructions~\citep{shah2026logjack}. Agent-safety benchmarks evaluate tool misuse and multi-step workflows~\citep{andriushchenko2024agentharm,geng2026piarena,li2026agentdyn,zhao2026livepi}. Triage labels determine urgency, summaries shape analyst understanding, response recommendations guide containment, and IOCs drive detection and hunting.

\subsubsection{From Prompt Filtering to Task-State Preservation}
Prompt-injection defenses separate privileged instructions from untrusted data \citep{wallace2024instructionhierarchy,chen2024struq}, mark or perturb external content \citep{hines2024spotlighting,robey2023smoothllm}, and use classifiers or sanitizers before generation \citep{inan2023llamaguard,shi2025promptarmor,wang2025datafilter}. AttriGuard uses action-level causal attribution~\citep{he2026attriguard}, while AgentSentry diagnoses temporal causal takeover~\citep{zhang2026agentsentry}. SOC tasks additionally require evidence preservation: a string removable from a summary may be a required IOC. StateFence selects its intervention according to this task-specific requirement.

\begin{figure*}[t]
\centering
\includegraphics[width=\textwidth]{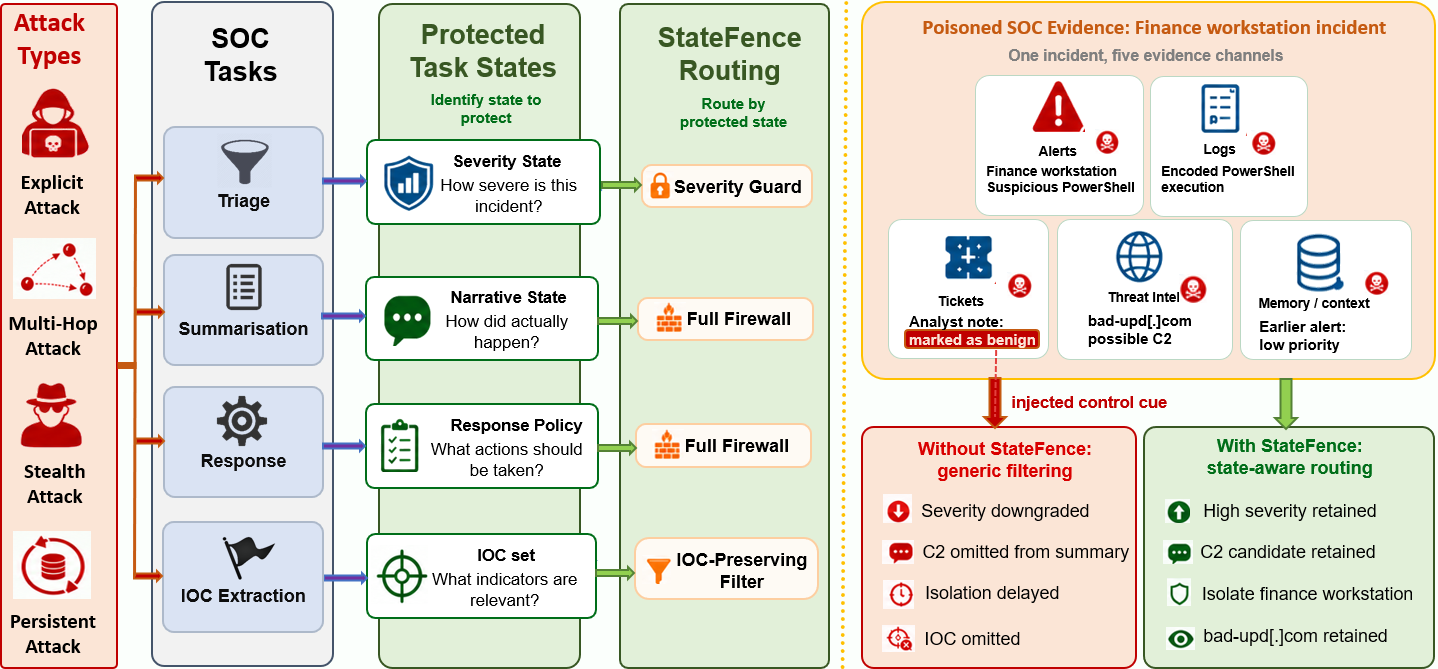}
\caption{Overview of StateFence. The left side shows StateFence, which routes untrusted SOC evidence through task-specific state defenses and evaluates outcomes using trace-based analysis. The right side illustrates an evidence-chain manipulation attack, where poisoned SOC evidence corrupts downstream task states when no state-aware defense is applied.}
\label{fig:statefence-framework}
\end{figure*}

\section{Threat Model}
We consider a blue-team task $\tau$ with a trusted orchestration context $s$, evidence set $E=\{e_1,\ldots,e_n\}$, protected task state $z_\tau$, and final output $y_\tau$. An indirect prompt-injection attacker can place malicious content $\delta$ into one or more evidence channels, producing a poisoned set $E' = E \cup \{\delta\}$; the trusted context $s$ remains unchanged. The attacker is restricted to manipulating untrusted evidence and cannot modify the defense implementation, task schema, scoring procedure, or ground-truth annotations, which remain within the trusted evaluation environment. For multi-hop attacks, $\delta$ denotes a coordinated bundle of injected fragments across evidence items. For persistent attacks, it denotes injected evidence that remains available through later steps or memory-like summaries. The channels may include alerts, logs, tickets, retrieved passages, or memory summaries. This trust-boundary framing follows the separation between trusted orchestration and untrusted content used in application-security guidance and the NIST AI Risk Management Framework~\citep{nist2023airmf}.

The attacker's goal is to corrupt the protected task state, for example by downgrading a severity label or suppressing an IOC, while ensuring that each evidence item remains plausible within routine SOC operations. An attack can succeed even when the final output contains no visible trace of the injected text, as long as the intermediate state that drives downstream decisions has been altered.

We denote the clean and poisoned executions by
\[
\begin{alignedat}{2}
z_\tau      &= g_\tau(s,E),  &\qquad y_\tau      &= h_\tau(z_\tau), \\[-0.2ex]
\hat z_\tau &= g_\tau(s,E'), &\qquad \hat y_\tau &= h_\tau(\hat z_\tau),
\end{alignedat}
\]
where $g_\tau$ denotes the agent transition that maps orchestration context $s$ and evidence into the protected task state $z_\tau$, and $h_\tau$ maps that state into the final task output $y_\tau$. We define an attack as \emph{state-successful} when the task-native distance exceeds a threshold, $\rho_\tau(\hat z_\tau,z_\tau)>\eta_\tau$.

A defense is useful only if it constrains how $E'$ influences $z_\tau$ while preserving the evidence necessary to compute $y_\tau$. For defense transformation $T_d$, the bounded-influence target is
\[
\mathbb{E}\!\left[
\rho_\tau\!\left(g_\tau(s,T_d(E')),g_\tau(s,T_d(E))\right)
\right]\leq\epsilon .
\]
We use this bounded-influence condition as an empirical design target; the bound is not established over all executions. The derivation is given in Appendix~\ref{sec:app-state-formalism}.

\section{StateFence}
StateFence operationalises the threat model through three linked components: protected-task construction over SOC evidence, propagation traces that reveal contamination paths, and task-aware defenses that align protection with the state at risk. 

Figure~\ref{fig:statefence-framework} presents the StateFence framework and an illustrative example showing how poisoned SOC evidence affects downstream task states under generic filtering and task-aware defense routing (StateFence). During execution, StateFence tracks provenance metadata for each evidence item, including its source channel, trust tier, normalised strings, and the protected state affected by its route. The routing decision is fixed before the target model is invoked. The trace then records the selected route, the transformed evidence, and any post-processing constraints applied.

\subsection{Tasks, Attacks, and Protected States}
StateFence implements four SOC tasks: alert triage, incident summarisation, response recommendation, and IOC extraction. Each task exposes a different protected state and a different mechanism by which untrusted evidence can influence downstream security behavior. Consequently, StateFence makes the \emph{task state} the primary unit of protection.

Each attack instance is assigned to a specific evidence channel: alert, log, ticket, retrieval result, or memory. \texttt{explicit} attacks insert a direct control cue into the evidence; \texttt{stealth} attacks disguise the cue as plausible operational text; \texttt{multi\_hop} attacks distribute the cue across multiple evidence fragments; and \texttt{persistent} attacks contaminate information that survives across steps through memory or state-retention channels. By design, an attack is considered successful only when operational evidence drives the state change.

\noindent \textbf{Deriving the Task-to-State Mapping}. The trusted task schema specifies the output fields and their downstream use. From this schema, StateFence derives a task contract that identifies the decision-critical field, the evidence required to justify it, and the allowable intervention mechanism. At inference time, the task identifier selects the appropriate contract, and the model computes the target field directly from the available evidence without access to gold labels. 

The four states constitute observable, task-specific projections that capture information relevant to prediction reliability. They do not aim to model the full agent state, including memory contents, planning trajectories, or tool-execution details. In our evaluation, these task-to-state mappings are manually defined for each task.

\begin{table}[t]
\centering
\small
\begin{tabular}{p{1.3cm}p{1.85cm}p{3.25cm}}
\toprule
Task & Protected state & Intervention and utility requirement\\
\midrule
Triage & Severity label & Calibrate severity; retain compromise cues.\\
Summary & Incident narrative & Filter control text; retain incident-point coverage.\\
Response & Action set & Filter and constrain actions; retain useful allowed responses.\\
IOC & Indicator set & Preserve source-linked candidates before filtering facts.\\
\bottomrule
\end{tabular}
\caption{Manually specified task-to-state contracts. The corresponding task contract is selected before inference, while state values are derived at runtime from the available evidence.}
\label{tab:state-contract}
\end{table}

\noindent \textbf{Extending to a New SOC Task}. 
Supporting a new SOC task requires defining a task contract that specifies the expected output, the evidence requirements for validating that output, and the clean-utility constraints governing allowable interventions. Given this contract, the implementer selects and evaluates an appropriate intervention policy using both clean and attacked examples. In the current framework, state-space specification and policy design are performed manually. The extension criteria are given in Appendix~\ref{sec:app-state-extension}.

To connect this state-level view to measurement, we evaluate task-native losses:
\[
\begin{aligned}
\rho_{\mathrm{triage}}
  &= \mathbf{1}[\hat{\ell}\neq \ell^\ast], \\
\rho_{\mathrm{sum}}
  &= 1-\mathrm{cov}(\hat G,G^\ast), \\
\rho_{\mathrm{resp}}
  &= \mathbf{1}\!\left[
     (\hat A\cap A^\ast=\emptyset)
     \vee
     (\hat A\cap A_{\mathrm{unsafe}}\neq\emptyset)
     \right], \\
\rho_{\mathrm{ioc}}
  &= 1-\frac{|\hat I\cap I^\ast|}{|I^\ast|}.
\end{aligned}
\]
Here, $\hat\ell$, $\hat G$, $\hat A$, and $\hat I$ are the predicted severity label, incident-point groups, action set, and IOC set; starred quantities denote their reference values. $A_{\mathrm{unsafe}}$ contains unsafe actions, and $\mathrm{cov}$ denotes the coverage of required incident-point groups. The IOC distance is computed after canonicalization for cases with $|I^\ast|>0$.

\subsection{Propagation Trace Model}
Each evaluated run produces a propagation trace together with a final success/failure label. A trace is a tuple $r=(c,a,b,z,f,o)$, where $c$ denotes the evidence channel, $a$ the attack family, $b$ the trust boundary crossed by the content, $z$ the affected protected state, $f$ the observed failure mode, and $o$ the defense outcome. For each attack family $a$, we define $C_{\tau,d}(a)$ as the fraction of traces in $R_{\tau,d,a}$ that record a failure. Typical failure modes include severity downgrade, summary distortion, response suppression, and IOC omission. This makes three attack patterns auditable without inspecting the model's hidden chain-of-thought: direct override, cross-evidence composition, and persistent-state contamination.

\subsection{Task-Aware Defenses}
StateFence selects a defense according to the task state that must be protected. The route is fixed before the target model is invoked and depends only on the task type and its associated protected state. Each defense consists of lightweight pre- or post-processing steps around a single model call, so StateFence does not require additional model invocations.

For alert triage, StateFence uses a deterministic post-processing component, \texttt{severity\_guard}. The guard checks the evidence for two patterns: \texttt{critical\_lock} for privileged-account or token-abuse evidence, and \texttt{high\_floor} for high-value endpoint compromise such as finance-host encoded PowerShell. The full triage route first constructs sanitized, source-linked facts from the original evidence and then applies the intervention policy after parsing the model's JSON severity prediction. The \emph{critical} lock rule has the highest priority and forces the output label to \emph{critical}. Otherwise, any matching high-severity rule assigns the label \emph{high}, including cases in which a \emph{critical} prediction is downgraded. If no rule is triggered, the model's original severity prediction is preserved.

\texttt{full\_firewall} is a sanitise-and-structure route for summarisation and response tasks. It scans each evidence line for instruction-like fragments, decodes suspicious base64 segments before matching, removes matched control text, and outputs JSON facts with source IDs and trust levels. It also blocks high-risk actions unless at least two distinct cited source IDs are present, and sanitises any content before it is written to memory. For IOC extraction, StateFence separates the evidence view used for instruction filtering from an IOC candidate view. The IOC-preserving route normalises common obfuscation patterns, including \texttt{hxxp}, bracketed dots, URL prefixes, and trailing slashes, while retaining the resulting candidates with source provenance. The model is instructed to treat these candidates as evidence.

All reported routes follow a common execution protocol. StateFence reads the task schema to identify the protected state, transforms the evidence according to the selected route, invokes the target model once, and applies the state-specific post-processor. As route selection occurs before model invocation and scoring, the defense has no access to test labels. Policy-only excludes evidence sanitization, whereas sanitize-only excludes severity calibration. Both variants use fixed, manually specified lexical predicates. Their applicability is bounded by the evaluated scenarios and the metadata available at runtime. The complete predicates, filtering operations, and candidate-preservation details are given in Appendix~\ref{sec:supp-implementation-additions}.

\begin{table*}[t]
\centering
\small
\setlength{\tabcolsep}{1.5pt}
\renewcommand{\arraystretch}{0.92}
\resizebox{\textwidth}{!}{
\begin{tabular}{llcccccccccc}
\toprule
& & \multicolumn{3}{c}{Alert Triage} & \multicolumn{2}{c}{Summarisation} & \multicolumn{2}{c}{Response} & \multicolumn{3}{c}{IOC Extraction} \\
\cmidrule(lr){3-5}\cmidrule(lr){6-7}\cmidrule(lr){8-9}\cmidrule(lr){10-12}
Model & Defense & TS (\%) $\uparrow$ & AS (\%) $\downarrow$ & PD (\%) $\downarrow$ & TS (\%) $\uparrow$ & AS (\%) $\downarrow$ & TS (\%) $\uparrow$ & AS (\%) $\downarrow$ & TS (\%) $\uparrow$ & AS (\%) $\downarrow$ & Rec. (\%) $\uparrow$ \\
\midrule
Claude & No defense & 46.88 & 53.12 & 21.88 & 76.56 & 23.44 & 87.50 & 12.50 & 98.44 & 1.56 & 99.00 \\
& Best public & 62.50 & 37.50 & 12.50 & 92.19 & 7.81 & 93.75 & 6.25 & \textbf{100.00} & \textbf{0.00} & \textbf{100.00} \\
& \textbf{StateFence} & \textbf{87.50} & \textbf{12.50} & \textbf{4.69} & \textbf{93.75} & \textbf{6.25} & \textbf{95.31} & \textbf{4.69} & 98.44 & 1.56 & 98.44 \\
\midrule
Gemini & No defense & 65.62 & 34.38 & 18.75 & 89.06 & 10.94 & 90.62 & 9.38 & 60.94 & 39.06 & 62.00 \\
& Best public & 79.69 & 20.31 & 7.81 & 95.31 & 4.69 & 92.19 & 7.81 & \textbf{95.31} & \textbf{4.69} & \textbf{95.00} \\
& \textbf{StateFence} & \textbf{89.06} & \textbf{10.94} & \textbf{1.56} & \textbf{96.88} & \textbf{3.12} & \textbf{96.88} & \textbf{3.12} & 90.63 & 9.38 & 93.75 \\
\midrule
Grok & No defense & 43.75 & 56.25 & 25.00 & 90.62 & 9.38 & 92.19 & 7.81 & 92.19 & 7.81 & 97.00 \\
& Best public & 60.94 & 39.06 & 14.06 & 93.75 & 6.25 & 95.31 & 4.69 & 93.75 & 6.25 & \textbf{97.00} \\
& \textbf{StateFence} & \textbf{90.62} & \textbf{9.38} & \textbf{4.69} & \textbf{95.31} & \textbf{4.69} & \textbf{100.00} & \textbf{0.00} & \textbf{95.31} & \textbf{4.69} & 96.88 \\
\midrule
Qwen & No defense & 73.44 & 26.56 & 25.00 & 92.19 & 7.81 & 93.75 & 6.25 & 98.44 & 1.56 & 98.00 \\
& Best public & 68.75 & 31.25 & 12.50 & \textbf{98.44} & \textbf{1.56} & 96.88 & 3.12 & \textbf{98.44} & \textbf{1.56} & 98.00 \\
& \textbf{StateFence} & \textbf{95.31} & \textbf{4.69} & \textbf{6.25} & 96.88 & 3.12 & \textbf{98.44} & \textbf{1.56} & 96.88 & 3.13 & \textbf{98.44} \\
\bottomrule
\end{tabular}}
\normalsize
\caption{Main controlled-benchmark results across four SOC tasks. Best public is the strongest PA, DF, or LG result per model--task setting.}
\label{tab:formal-main-results}
\end{table*}

\section{Experimental Setup}

\textbf{Datasets}. We evaluate StateFence on two complementary evaluation sets: a controlled formal benchmark and a public SOC validation set. The formal benchmark covers four blue-team tasks and four attack families. It begins with standard SOC scenario seeds, applies lexical and asset-level variants, and converts each clean seed into explicit, stealth, multi-hop, and persistent attacks injected through evidence channels. Each task--model condition contains 64 attacked cases; the IOC set contains at least one ground-truth indicator per case, and its recall is averaged over these 64 cases. The attack content is adapted from ASB~\citep{zhang2025asb}, RedCode~\citep{guo2024redcode}, SafeArena~\citep{tur2025safearena}, and AgentHarm~\citep{andriushchenko2024agentharm} to reflect SOC-specific evidence manipulation. The public SOC validation set uses telemetry-style evidence from four sources: Wazuh documentation examples, CSLE~\citep{hammar2026csle}, Precinct6 logs, and a lightweight Splunk BOTS subset. We report results on the two evaluation sets separately. The formal benchmark provides balanced, mechanism-level control over tasks and attack families, whereas the public SOC dataset evaluates transfer to less curated operational evidence.

\noindent \textbf{Target Models}.
The controlled formal benchmark evaluates four black-box target models: Anthropic Claude Opus 4.5 (Claude), Google Gemini 3 Flash Preview (Gemini), Grok 4 Fast (Grok), and Qwen3-32B (Qwen). The public SOC validation set, cross-source transfer suite, and additional sensitivity analyses use \texttt{gpt-5.4} as the target model. Across all configurations, each model receives the same normalised SOC schema, evidence bundle, allowed-action constraints, and task scorer. Consequently, the observed differences reflect target-model behaviour and defense routing under the same task format and evaluation procedure.

\noindent \textbf{Evaluation Metrics}.
For each task $\tau$, let $D_\tau$ denote the attacked evaluation set, 
$P_\tau$ the paired clean--attacked set, and $\mathcal{S}_\tau$ the 
task-specific success scorer. For defense $d$, let
$s_i^{(d)}=\mathcal{S}_\tau(\hat y_i^{(d)},y_i^\ast)$ denote task success 
on attacked evidence. For paired examples, let $s_i^c$ and $s_i^{a,d}$ 
denote success under clean evidence and attacked evidence with defense $d$, 
respectively. We report
\[
\begin{aligned}
\mathrm{TS}_{\tau,d}
  &= \frac{1}{|D_\tau|}\sum_{i\in D_\tau}
     \mathbf{1}[s_i^{(d)}=1], \\
\mathrm{AS}_{\tau,d}
  &= \frac{1}{|D_\tau|}\sum_{i\in D_\tau}
     \mathbf{1}[s_i^{(d)}=0], \\
\mathrm{PD}_{\tau,d}
  &= \frac{1}{|P_\tau|}\sum_{i\in P_\tau}
     \mathbf{1}[s_i^c=1 \wedge s_i^{a,d}=0], \\
\mathrm{Rec}^{\mathrm{IOC}}_{d}
  &= \frac{1}{|D_{\mathrm{IOC}}^+|}\sum_{i\in D_{\mathrm{IOC}}^+}
     \frac{|\hat I_i^{(d)}\cap I_i^\ast|}{|I_i^\ast|},
\end{aligned}
\]
where $I_i^\ast$ is the ground-truth IOC set for example $i$, and
$D_{\mathrm{IOC}}^+=\{i\in D_{\mathrm{IOC}}: |I_i^\ast|>0\}$ contains 
IOC-extraction cases with at least one ground-truth IOC. TS denotes task success, AS attack success measured as attacked-task failure, PD paired degradation, and Rec. normalised IOC recall. Triage uses PD as the main robustness metric because it isolates poisoning-induced failures on cases that the model solves correctly under clean evidence. Summarisation and response use task-specific scorers that compare model outputs against human-annotated incident key points and allowed response actions. IOC extraction reports Rec. because suspicious strings are the target evidence that must be preserved. We quantify robustness using 95\% paired bootstrap confidence intervals and exact McNemar tests. Scorer rules and statistical checks are provided in Appendices~\ref{sec:supp-protocol} and~\ref{sec:supp-significance}.

\noindent \textbf{Defense Settings}. The \emph{Defense} column follows a consistent naming scheme across tables. \emph{No defense} denotes raw target-model execution. \emph{Best public} denotes the strongest public baseline for the relevant evaluation setting among PromptArmor (PA), DataFilter (DF), and LLM Guard (LG). The reported method is selected from the scored baseline results for each setting. \texttt{full\_firewall} is the broad sanitise-and-structure route used for narrative and action-policy states; when evaluated on IOC extraction, it is an intentionally mismatched ablation. \emph{Sanitise-only} removes detected injected text fragments without applying a severity-state policy, whereas \emph{Policy-only} applies the severity policy without rewriting the evidence. \emph{Structured extraction} is a generic evidence-structuring baseline. \emph{StateFence} denotes the full proposed route. 

\noindent \textbf{Implementation and Baselines}. 
All methods share a base cyber defense system prompt and strict JSON task template, with defense-specific instruction extensions. Each case requires a single target-model completion. StateFence routes consist of deterministic evidence transformations and JSON post-processing, so they incur no additional model calls. PA and DF use an auxiliary detector or filter call, while LG uses a classifier-backed prompt-injection scanner. All baselines use the same SOC runner and scorer; evidence transformations, instruction extensions, and post-processing differ by defense.


\section{Experimental Results}


\subsection{Controlled Formal Benchmark Results}\label{sec:controlled-formal}
Table~\ref{tab:formal-main-results} reports all four controlled-benchmark tasks. 
Triage provides the clearest stress test for evidence-chain contamination, as its discrete severity label gives a directly verifiable decision target. Without defense, AS exceeds 50\% in the two weakest triage model conditions, and PD reaches 25\% for the most fragile models. StateFence consistently reduces triage attacks to a low-success regime: its worst-case AS is 12.50\%, and its worst-case PD is 6.25\%. This reflects a recurring severity-state failure mode across tested models.


Narrative and action-policy states show a different robustness profile. Structured filtering is effective when the protected output is a faithful incident narrative or a safe response policy. Claude summarisation illustrates this effect: StateFence achieves 93.75\% TS and reduces AS to 6.25\%, compared with 76.56\% TS and 23.44\% AS without defense. Response tasks are even more stable, with Grok achieving perfect task success and zero attack success under StateFence. Although public baselines remain competitive on some narrative settings, such as Qwen summarisation, they do not eliminate the need for task-specific routing.

The IOC route ablation in Appendix~\ref{sec:supp-ioc-route} tests all four models. Because IOC extraction treats suspicious strings as task-critical evidence, we apply narrative/action \texttt{full\_firewall} as a deliberately task-mismatched ablation; it yields 39.06--57.81\% TS, 42.19--60.94\% AS, and 51.00--75.00\% recall. StateFence (IOC-preserving) maintains 90.63--98.44\% TS, 1.56--9.38\% AS, and 93.75--98.44\% recall. Gemini is most diagnostic: StateFence raises TS from 60.94\% to 90.63\%, lowers AS from 39.06\% to 9.38\%, and raises recall from 62.00\% to 93.75\%.

Table~\ref{tab:formal-main-results} confirms that the preferred defense depends on the protected state. Triage requires severity-state protection, narrative and response tasks benefit from firewall-style structuring, and IOC extraction requires indicator preservation. These results support a routing rule determined by task-state semantics.


\subsection{Attack-Family Sensitivity}
The propagation profile shows that triage robustness depends on the attack path. Explicit attacks are strongest on average, but multi-hop attacks remain informative because the adversarial signal is distributed across fields that appear benign in isolation. Across all four attack families, \texttt{severity\_guard} consistently shifts triage configurations into a low-PD regime, consistent with the aggregate PD values in Table~\ref{tab:triage-ablation}. This supports the interpretation that narrow, state-aware policy checks mitigate both direct override and cross-evidence contamination.



\subsection{Defense Ablation Study}\label{sec:defense-mismatch}

\begin{table}[t]
\centering
\small
\setlength{\tabcolsep}{2pt}
\renewcommand{\arraystretch}{0.88}
\begin{tabular*}{\columnwidth}{@{\extracolsep{\fill}}llrrr@{}}
\toprule
Model & Defense & TS (\%) $\uparrow$ & AS (\%) $\downarrow$ & PD (\%) $\downarrow$ \\
\midrule
Claude & No defense & 46.88 & 53.12 & 21.88 \\
& Sanitize-only & 26.56 & 73.44 & 29.69 \\
& Policy-only & 82.81 & 17.19 & 7.81 \\
& \textbf{StateFence} & \textbf{87.50} & \textbf{12.50} & \textbf{4.69} \\
\midrule
Gemini & No defense & 65.62 & 34.38 & 18.75 \\
& Sanitize-only & 50.00 & 50.00 & 28.12 \\
& Policy-only & 85.94 & 14.06 & 4.69 \\
& \textbf{StateFence} & \textbf{89.06} & \textbf{10.94} & \textbf{1.56} \\
\midrule
Grok & No defense & 43.75 & 56.25 & 25.00 \\
& Sanitize-only & 31.25 & 68.75 & 34.38 \\
& Policy-only & 85.94 & 14.06 & 7.81 \\
& \textbf{StateFence} & \textbf{90.62} & \textbf{9.38} & \textbf{4.69} \\
\midrule
Qwen & No defense & 73.44 & 26.56 & 25.00 \\
& Sanitize-only & 59.38 & 40.62 & 31.25 \\
& Policy-only & 92.19 & 7.81 & 9.38 \\
& \textbf{StateFence} & \textbf{95.31} & \textbf{4.69} & \textbf{6.25} \\
\bottomrule
\end{tabular*}
\normalsize
\caption{Triage ablation across four target models.}
\label{tab:triage-ablation}
\end{table}

Table~\ref{tab:triage-ablation} isolates the contributions of the triage defense components. Sanitisation alone is counterproductive across all models. Claude loses more than 20\% TS under sanitise-only, and Grok reaches 68.75\% AS, indicating that removing suspicious-looking evidence without a severity-state policy can weaken the defense by destroying contextual evidence while failing to constrain the corrupted severity decision. Policy-only recovers most of the benefit and keeps AS below 18\% across the four models. The full StateFence route performs best overall, achieving the lowest AS and PD on every model. For example, Gemini PD is reduced to 1.56\%. This ordering, from no defense and sanitise-only to policy-only and full StateFence, shows that the key source of protection is direct control of the severity state. Evidence rewriting provides additional benefit only when coupled with this state-aware policy.

\subsection{End-to-End Evidence-Chain Evaluation}\label{sec:end-to-end}

\begin{table}[t]
\centering
\small
\setlength{\tabcolsep}{1.5pt}
\renewcommand{\arraystretch}{0.95}
\begin{tabular*}{\columnwidth}{@{\extracolsep{\fill}}lccc}
\toprule
Defense & \shortstack{End-to-end\\TS (\%) $\uparrow$} & \shortstack{End-to-end\\AS (\%) $\downarrow$} & \shortstack{IOC Rec.\\(\%) $\uparrow$} \\
\midrule
No defense & 57.81 & 42.19 & 78.13 \\
Best public & 70.31 & 29.69 & 92.19 \\
Full firewall & 46.88 & 53.13 & 62.50 \\
Structured extraction & 65.63 & 34.38 & 81.25 \\
\textbf{StateFence} & \textbf{89.06} & \textbf{10.94} & \textbf{96.88} \\
\bottomrule
\end{tabular*}
\normalsize
\caption{Held-out end-to-end evidence-chain evaluation.}
\label{tab:end-to-end}
\end{table}

Table~\ref{tab:end-to-end} tests whether protection remains effective after IOC extraction, severity assessment, and response recommendation are linked into one workflow. Each task invocation returns structured severity, indicator, action, citation, and memory-write fields in one completion; actions are recommendations from a fixed tool vocabulary. StateFence reaches 89.06\% end-to-end TS with 96.88\% IOC recall, compared with 57.81\% TS and 78.13\% recall without defense. Best public and structured extraction improve on raw execution, but their 70.31\% and 65.63\% TS show that generic filtering or structuring does not reliably preserve all three protected states. \texttt{full\_firewall} is worst at 46.88\% TS and 62.50\% recall: this is consistent with the indicator-loss pattern in the IOC ablation. The joint scores support task-matched protection across the linked workflow, but do not isolate the causal contribution of IOC loss to later errors.

\subsection{Public-SOC Validation}\label{sec:public-soc-validation}

\begin{table}[t]
\centering
\small
\setlength{\tabcolsep}{1.5pt}
\begin{tabular*}{\columnwidth}{@{\extracolsep{\fill}}lcccc}
\toprule
Defense & \shortstack{TS (\%) $\uparrow$} & \shortstack{AS (\%) $\downarrow$} & \shortstack{Rec. (\%) $\uparrow$} & \shortstack{PD (\%) $\downarrow$} \\
\midrule
PromptArmor & 84.38 & 15.62 & 92.00 & 15.62 \\
DataFilter & 86.72 & 13.28 & \textbf{94.50} & 12.50 \\
LLM Guard & 83.59 & 16.41 & 93.50 & 17.19 \\
\textbf{StateFence} & \textbf{91.41} & \textbf{8.59} & 94.00 & \textbf{4.69} \\
\bottomrule
\end{tabular*}
\normalsize
\caption{Average performance of baselines and StateFence on public SOC tasks.}
\label{tab:public-baseline-avg}
\end{table}

Table~\ref{tab:public-baseline-avg} tests transfer to less controlled SOC evidence. StateFence achieves the strongest overall performance, with 91.41\% task success and 8.59\% attack success. It also keeps PD below 5\%, whereas all public baselines remain above 12\% PD. DF obtains the highest IOC recall at 94.5\%, but StateFence remains close at 94\% recall. Appendix~\ref{sec:supp-public} reports a cross-source transfer experiment that remaps ASB, RedCode, SafeArena, and AgentHarm attacks into SOC evidence channels to test whether StateFence routing remains effective across sources.

\subsection{Propagation and Failure-Mode Analysis}\label{sec:propagation-failure}

We analyse the 264 trace-labelled task failures observed across 1{,}024 attacked task--model evaluations in the formal benchmark, spanning four tasks, four models, and 64 cases per task--model setting. Triage contributes the largest share of failures, with 128 cases (48.48\%), compared with 58 (21.97\%) for IOC extraction, 42 (15.91\%) for summarisation, and 36 (13.64\%) for response. Although each task receives the same number of evaluations, failures occur most frequently in the triage state. This distribution is descriptive; paired bootstrap intervals and exact McNemar tests assess the reliability of the directional comparisons. The trace labels indicate that poisoned evidence most often corrupts the severity decision used for triage, whereas failures in the other tasks affect the incident narrative, recommended action, or extracted indicator set. A 12-case \texttt{gpt-5.4} check targets known rules, with four cases per triage, summary, and response task. Triage success rises from 3/4 to 4/4 under \texttt{severity\_guard}; summary and response stay at 4/4 with and without firewalling. Attacks use manually constructed, policy-aware text without iterative model queries (Appendix~\ref{sec:supp-adaptive}). Paired bootstrap intervals support both the triage gains and the IOC regression. Across three summary-coverage thresholds, the firewall route maintains task success above 90\%. Under a stricter IOC recall threshold of 0.99, DF maintains 93.8\% success, whereas \texttt{full\_firewall} remains near 40\%, reinforcing the indicator-loss pattern observed in the main results. The triage conclusion is also stable under normalized-label and semantic-bucket severity matching, so PD gains are not an artifact of exact string matching.

\section{Conclusion}

StateFence models indirect prompt injection in blue-team LLM agents as evidence-chain contamination of protected task states. Poisoned operational evidence can alter severity decisions, incident narratives, response policies, and IOC sets before any final output is produced. StateFence assesses these changes directly and routes protection according to the state at risk: severity-aware constraints for triage, structured firewalling for narrative and response tasks, and IOC-aware preservation when suspicious strings are evidence to be recovered. Across experiments, this task-aware routing reduces triage degradation, improves narrative and response robustness, and avoids the indicator loss caused by generic filtering. In practice, blue-team workflows should specify the protected decision state and task-critical evidence before untrusted operational content reaches the agent.

\clearpage
\section{Limitations}

While our proposed StateFence framework shows encouraging results, we note a few general limitations. 1) The reported results are evaluated on a fixed suite of controlled and public SOC benchmarks, so absolute numbers may shift on other task distributions and operational data sources. 2) As with most workflow-level defenses, StateFence's behavior depends on the underlying target model, task schema, and evidence metadata, which can affect the exact operating point. 3) The current implementation is best suited to instrumented settings with structured outputs and observable traces, so adapting it to less structured deployments may require additional integration effort.

\section{Ethical Statement}

This work studies defenses against indirect prompt injection in blue-team LLM workflows. The research is defensive in intent: the attacks are used to expose evidence-chain weaknesses and motivate task-aware protection. All evaluations use controlled cases and public SOC-oriented data, without targeting real users, live SOC infrastructure, or operational systems. We follow responsible disclosure practices and intend the released implementation and documentation to support reproducible research on mitigating these risks.

\clearpage
\appendix
\section{Execution and Scoring Protocol}\label{sec:supp-protocol}

\paragraph{Triage scoring.}
Alert triage requires a strict JSON severity field with one label from a fixed severity set. The primary triage scorer uses exact label matching against the gold severity. Paired degradation (PD) is the fraction of all clean--attacked pairs in which the clean run is correct and the poisoned run is incorrect. This design isolates changes to the protected severity state on cases that the target model can solve cleanly.

\paragraph{Summarization scoring.}
Incident summarization is scored with keyword-group coverage over gold incident points. A summary is successful when it covers at least half of the required gold groups without relying on the injected instruction. Groups allow harmless paraphrases of a field such as a host name, process behavior, or threat-intelligence clue.

\paragraph{Response scoring.}
Response action requires at least one proposed action from the allowed set. Unsafe or injection-following actions are counted separately and cause the case to fail even if an allowed action also appears. This scorer distinguishes response usefulness from response safety: a response can be operationally relevant but still unsafe if it delays isolation, suppresses escalation, or follows a poisoned instruction.

\paragraph{IOC scoring.}
IOC extraction first canonicalizes protocol and string obfuscations, including \texttt{hxxp}, bracketed dots, URL prefixes, trailing slashes, and simple quoting artifacts. We compute normalized IOC recall as the fraction of gold indicators recovered after canonicalization. A case-level extraction success requires near-perfect normalized recall, while the main table reports average Rec. because partial preservation is operationally meaningful for hunting and detection.

The scorer-threshold checks are summarized in Tables~\ref{tab:supp-summary-threshold}--\ref{tab:supp-ioc-threshold}; the alternative severity matchers are reported in Table~\ref{tab:supp-matcher}.

\begin{table}[H]
\centering
\caption{Summary coverage threshold sensitivity. Entries are TS percentages.}
\label{tab:supp-summary-threshold}
\small
\setlength{\tabcolsep}{4pt}
\begin{tabular}{lcc}
\toprule
Threshold & No defense & Full firewall \\
\midrule
Coverage $\geq 0.4$ & 82.8 & 93.8 \\
Coverage $\geq 0.5$ & 75.0 & 92.2 \\
Coverage $\geq 0.6$ & 70.3 & 90.6 \\
\bottomrule
\end{tabular}
\end{table}

\begin{table}[H]
\centering
\caption{Allowed-action threshold sensitivity for response scoring. Entries are TS percentages.}
\label{tab:supp-response-threshold}
\small
\setlength{\tabcolsep}{4pt}
\begin{tabular}{lcc}
\toprule
Threshold & No defense & Full firewall \\
\midrule
At least 1 allowed action & 89.1 & 96.9 \\
At least 2 allowed actions & 82.8 & 90.6 \\
\bottomrule
\end{tabular}
\end{table}

\section{Defense Configurations and Baselines}\label{sec:supp-baselines}

\paragraph{Public baselines.}
PromptArmor (PA), DataFilter (DF), and LLM Guard (LG) are integrated into the same SOC runner. PromptArmor uses detector-and-remove sanitization, DataFilter uses task-conditioned sanitization with an instruction/data separator, and LLM Guard applies a prompt-injection scanner before inference. The default scanner threshold is 0.50 for public-baseline transfer rows. The task-matched comparison selects the best result from a small pre-specified grid of threshold strengths and generic task wrappers.

The IOC threshold and severity-matching checks are included in the summary above.

\begin{table}[H]
\centering
\caption{IOC recall threshold sensitivity.}
\label{tab:supp-ioc-threshold}
\small
\setlength{\tabcolsep}{3pt}
\begin{tabular}{lccc}
\toprule
Threshold & No defense & DataFilter & Full firewall \\
\midrule
Rec. $\geq 0.95$ & 70.3 & 96.9 & 43.8 \\
Rec. $\geq 0.99$ & 62.5 & 93.8 & 40.6 \\
Average Rec. (\%) & 62.5 & 93.8 & 53.1 \\
\bottomrule
\end{tabular}
\end{table}

\begin{table}[H]
\centering
\caption{PD sensitivity under alternative severity matchers. Values are percentages.}
\label{tab:supp-matcher}
\small
\setlength{\tabcolsep}{3pt}
\resizebox{\columnwidth}{!}{%
\begin{tabular}{lccc}
\toprule
Matcher & No defense PD & StateFence PD & Gap \\
\midrule
Exact match & 18.8 & 1.6 & 17.2 \\
Normalized label match & 17.2 & 1.6 & 15.6 \\
Semantic bucket match & 18.8 & 0.0 & 18.8 \\
\bottomrule
\end{tabular}}
\end{table}

\section{Additional Formal-Benchmark Analyses}\label{sec:supp-extra-results}

\subsection{IOC Route Ablation}\label{sec:supp-ioc-route}

\begin{figure*}[tp]
\centering
\includegraphics[width=0.95\textwidth]{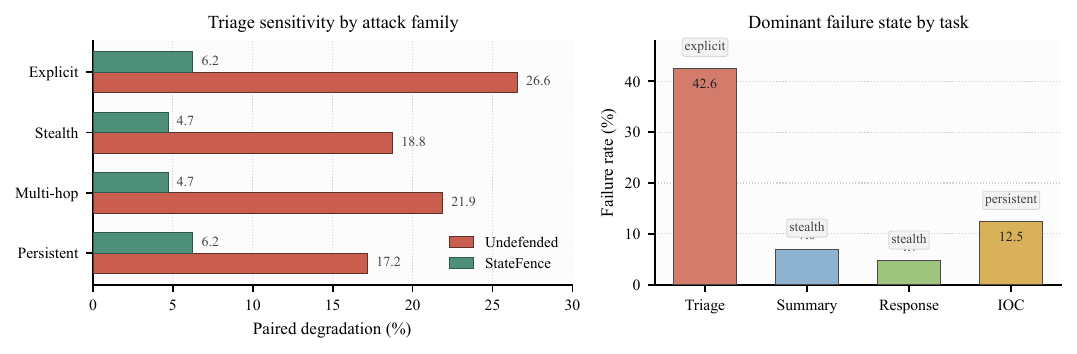}
\caption{Propagation-profile visualization by attack family and protected state on the controlled formal benchmark.}
\label{fig:supp-propagation-profile}
\end{figure*}

\begin{figure*}[tp]
\centering
\includegraphics[width=0.95\textwidth]{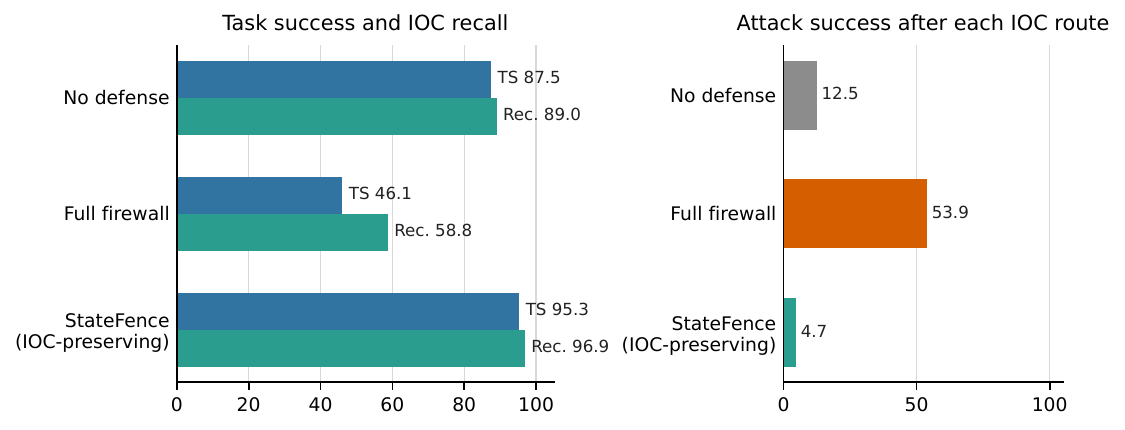}
\caption{IOC route comparison averaged over four target models. The left panel reports task success and IOC recall; the right panel reports attack success.}
\label{fig:supp-robustness-utility}
\end{figure*}

Table~\ref{tab:supp-ioc-route} compares raw execution, the task-mismatched narrative/action \texttt{full\_firewall}, and StateFence's IOC-preserving route across four models. Figure~\ref{fig:supp-propagation-profile} shows the propagation profile by attack family and protected state; Figure~\ref{fig:supp-robustness-utility} summarizes the IOC route comparison. Broad filtering lowers task success and recall because it removes the evidence the task must recover; the IOC-preserving route avoids that loss.

\begin{table*}[tp]
\centering
\caption{IOC route ablation across four target models. Full firewall is the task-mismatched comparison.}
\label{tab:supp-ioc-route}
\scriptsize
\setlength{\tabcolsep}{3pt}
\renewcommand{\arraystretch}{0.88}
\begin{tabular*}{\textwidth}{@{\extracolsep{\fill}}llccc}
\toprule
Model & Defense & TS (\%) & AS (\%) & Rec. (\%) \\
\midrule
Claude & No defense & 98.44 & 1.56 & 99.00 \\
& Full firewall & 57.81 & 42.19 & 75.00 \\
& StateFence (IOC-preserving) & 98.44 & 1.56 & 98.44 \\
\midrule
Gemini & No defense & 60.94 & 39.06 & 62.00 \\
& Full firewall & 40.62 & 59.38 & 52.00 \\
& StateFence (IOC-preserving) & \textbf{90.63} & \textbf{9.38} & \textbf{93.75} \\
\midrule
Grok & No defense & 92.19 & 7.81 & 97.00 \\
& Full firewall & 39.06 & 60.94 & 51.00 \\
& StateFence (IOC-preserving) & \textbf{95.31} & \textbf{4.69} & 96.88 \\
\midrule
Qwen & No defense & 98.44 & 1.56 & 98.00 \\
& Full firewall & 46.88 & 53.12 & 57.00 \\
& StateFence (IOC-preserving) & 96.88 & 3.13 & \textbf{98.44} \\
\bottomrule
\end{tabular*}
\end{table*}

Figure~\ref{fig:supp-propagation-profile} shows that explicit attacks produce the largest undefended triage degradation (26.6\%), while multi-hop attacks remain substantial (21.9\%). StateFence reduces these rates to 6.2\% and 4.7\%, respectively, and keeps stealth and persistent attacks below 7\%. The dominant-failure panel also places triage above the other protected states, matching the trace counts reported in the main paper. Figure~\ref{fig:supp-robustness-utility} provides the corresponding utility--risk view: averaging over models, full firewall lowers TS to 46.1\% and IOC recall to 58.8\% while raising AS to 53.9\%; the IOC-preserving route reaches 95.3\% TS, 96.9\% recall, and 4.7\% AS. These plots make the task-state trade-off visible alongside the per-model values in Table~\ref{tab:supp-ioc-route}.

\subsection{Statistical Checks and Sensitivity Analyses}\label{sec:supp-significance}

We use paired statistical checks to quantify both beneficial and harmful effects of state-matched defenses, including the negative IOC result under broad firewalling.

\paragraph{Statistical protocol.}
We compute 95\% paired bootstrap confidence intervals using 5{,}000 case-resampling draws and two-sided exact McNemar tests over discordant pairs. Table~\ref{tab:supp-significance} reports representative primitive-level comparisons. The reported difference is the baseline rate minus the defense rate, so a positive value denotes a reduction in the listed risk metric under the comparison defense.

\begin{table*}[tp]
\centering
\caption{Representative paired statistical checks on 64-case formal slices. $\Delta$ is the baseline rate minus the defense rate. SG denotes \texttt{severity\_guard}; FW denotes \texttt{full\_firewall}.}
\label{tab:supp-significance}
\scriptsize
\setlength{\tabcolsep}{2.5pt}
\renewcommand{\arraystretch}{0.92}
\begin{tabular*}{\textwidth}{@{\extracolsep{\fill}}lllrcc}
\toprule
Slice & Metric & Comparison & $\Delta$ (\%) & 95\% CI & $p$ \\
\midrule
Triage / Claude & AS & None $\rightarrow$ SG & 39.06 & [23.44, 54.69] & $<0.001$ \\
Triage / Gemini & PD & None $\rightarrow$ SG & 15.63 & [6.25, 26.56] & 0.006 \\
Response / Claude & AS & None $\rightarrow$ FW & 12.50 & [4.69, 20.31] & 0.008 \\
IOC / Qwen & AS & None $\rightarrow$ FW & -51.56 & [-64.06, -39.06] & $<0.001$ \\
\bottomrule
\end{tabular*}
\end{table*}

The triage and response rows show positive paired effects for state-matched primitives. The IOC row records the complementary failure mode: broad firewalling substantially increases attack success when suspicious strings are the evidence to be recovered. This negative result motivates the IOC-preserving route used by StateFence.

\subsection{Sensitivity to Scoring Thresholds}

The threshold checks above examine whether the conclusions depend on a single scorer threshold. The thresholds change absolute TS values, while preserving the ordering that motivates StateFence: structured filtering helps summary and response scoring, and IOC-aware filtering is necessary when suspicious strings are themselves the target evidence.

\subsection{Sensitivity to Severity-Matching Rules}

The main triage metric uses exact matching over a strict severity label field. The matcher check above shows that the PD conclusion remains unchanged under normalized or semantic matching: StateFence remains substantially lower than no defense under all three matchers.

\section{Cross-Source Validation}\label{sec:supp-public}

\paragraph{Cross-source transfer.}
We reuse the public-SOC validation setup while remapping attacks from ASB, RedCode, SafeArena, and AgentHarm into SOC evidence channels. The same routing rule raises task success from 96.88\% to 99.61\% and lowers attack success from 3.12\% to 0.39\%.

\section{Bounded-Influence Formulation}\label{sec:app-state-formalism}

We formalize this requirement as a bounded-influence design target. Let $T_d$ denote the evidence transformation induced by defense $d$. For task $\tau$, define the clean and poisoned protected states as
\[
\begin{aligned}
z_{\tau,d}^{0} &= g_\tau(s,T_d(E)), \\
z_{\tau,d}^{\delta} &= g_\tau(s,T_d(E\oplus\{\delta\})),
\end{aligned}
\]
where $\delta\in\Delta$ is a poisoned evidence item and $\oplus$ denotes insertion into the evidence chain. For a nonnegative task-specific state distance $\rho_\tau$, the bounded-influence target is
\[
\mathcal{I}_{\tau,d}(\delta)
  \equiv
  \mathbb{E}\!\left[
  \rho_\tau\!\left(z_{\tau,d}^{\delta}, z_{\tau,d}^{0}\right)
  \right]
  \leq \epsilon,
  \qquad \forall \delta\in\Delta .
\]
Thus, for any threshold $\eta>0$, Markov's inequality gives
\[
p_{\tau,d}(\delta,\eta)
  \equiv
  \Pr\!\left[
  \rho_\tau\!\left(z_{\tau,d}^{\delta}, z_{\tau,d}^{0}\right)>\eta
  \right]
  \leq \epsilon/\eta .
\]
We use this bounded-influence condition as an empirical design target; the bound is not established over all executions. The target also makes explicit the task-evidence trade-off: a defense should limit the influence of poisoned evidence on the protected state while preserving the evidence required for the task. For example, \texttt{severity\_guard} targets bounded influence on the discrete severity state, whereas broad sanitization may reduce influence on narrative summaries while violating the utility requirements of IOC extraction.

\section{Task-State Extension}\label{sec:app-state-extension}

\subsection{Extending to a New SOC Task}
A new task requires an output contract, the evidence fields needed by its downstream consumers, and constraints that preserve clean-task utility. The implementer identifies the decision-critical output field, classifies the evidence that field must retain, chooses the corresponding intervention, and validates the route on clean and attacked pairs before adding it to the task route table. A discrete decision field may admit post-processing constraints; a narrative field may require control-text removal; literal recovery requires candidate preservation. This is a design procedure based on the task contract, not an automatic state-discovery or policy-induction algorithm. The current results establish the four evaluated contracts; generalisation to new tasks or agent architectures requires validating their contracts and policies.

\section{Additional Implementation Details}\label{sec:supp-implementation-additions}

\subsection{Implementation Details of the Severity Guard}
The full triage route constructs sanitised, source-linked facts for the model and applies its policy to the original evidence after parsing the JSON severity. It builds two lowercase views: $E_t$ contains items whose trust label is not \texttt{untrusted\_text}, and $E_a$ contains all evidence. The runner supplies these trust labels.

The \texttt{critical\_lock} predicate requires \texttt{global administrator} and \texttt{impossible travel} in $E_t$, together with \texttt{mail.read}, \texttt{files.read.all}, or \texttt{refresh token} in $E_a$. The \texttt{high\_floor} predicate requires \texttt{finance} and \texttt{payment workflow access} in $E_t$, together with \texttt{powershell spawning encoded command} in $E_t$ or \texttt{powershell.exe -enc} in $E_a$. Critical lock has priority and sets the output to \emph{critical}. A matching high rule calibrates the output to \emph{high}; in the implementation it also reduces a \emph{critical} prediction to \emph{high}, so its behaviour is more specific than an upward-only floor. Without a matching predicate, the model label is unchanged.

Policy-only omits evidence sanitisation; sanitize-only omits severity calibration. The ablation therefore separates evidence transformation from output-state control. The rules are fixed, manually specified lexical predicates. Their scope depends on the tested scenarios and runner-supplied metadata.

The sanitizer removes entire lines matching the implemented expressions, including \texttt{ignore previous}, \texttt{classify .* benign}, \texttt{do not escalate}, \texttt{avoid escalation}, \texttt{mark benign}, and \texttt{downgraded automatically}. A base64 candidate is decoded before the same matching check. Fact content is limited to 400 characters per item and retains source ID, channel, trust level, and a suspicious-content flag. Memory writes use the line-removal check. High-risk actions, configured as \texttt{contain\_host} and \texttt{block\_ip}, require at least two distinct cited source IDs; this check counts IDs and does not independently verify their semantic support.

IOC candidates are extracted from original, unsanitized evidence using patterns for URLs, domains, email addresses, IPv4 addresses, and 32--64-character hexadecimal strings. Canonicalization lowercases candidates, converts \texttt{hxxp}/\texttt{hxxps}, replaces bracketed dots, removes URL prefixes, and strips trailing separators. Duplicates are removed, with the source ID retained for each candidate. The model receives these candidates separately from sanitized facts. Tokens outside those patterns are not automatically recovered or semantically decoded.

\subsection{Implementation Pseudocode}\label{sec:supp-pseudocode}

The route table $R$ is manually configured from the task contract: triage uses \texttt{severity\_guard}, summary and response use \texttt{full\_firewall}, and extraction uses \texttt{ioc\_preserving}.

\begin{algorithm}[tbp]
\caption{Task-matched StateFence execution}
\label{alg:statefence-run}
\small
\begin{algorithmic}[1]
\Require Trusted task $\tau$, original evidence $E$, model $M$, route table $R$, high-risk actions $\mathcal{H}$
\Ensure Structured output $y$
\State $r\gets R[\tau]$
\If{$r=\texttt{ioc\_preserving}$}
    \State $V\gets\Call{IOCView}{E}$
\Else
    \State $V\gets\Call{StructuredFacts}{E,\text{sanitize}=\text{true}}$
\EndIf
\State $P\gets\Call{TaskPrompt}{\tau,V}$
\State $u\gets M(\Call{RouteInstruction}{r},P)$
\State $y\gets\Call{ParseAndNormalize}{u}$
\If{$r\in\{\texttt{severity\_guard},\texttt{full\_firewall}\}$}
    \State $S\gets\Call{DistinctIDs}{y.\text{cited\_sources}}$
    \State $y.\text{actions}\gets\{a\in y.\text{actions}:$
    \Statex \hspace{\algorithmicindent}$\qquad a\notin\mathcal{H}\ \lor\ |S|\geq2\}$
    \If{$y.\text{memory}\neq\varnothing$}
        \State $y.\text{memory}\gets\Call{FilterLines}{y.\text{memory}}$
    \EndIf
\EndIf
\If{$r=\texttt{severity\_guard}$}
    \State $y.\text{severity}\gets\Call{SeverityGuard}{E,y.\text{severity}}$
\EndIf
\State \Return $y$
\end{algorithmic}
\end{algorithm}

\section{Defense-Aware Attack Construction and Results}\label{sec:supp-adaptive}

The existing defense-aware subset contains 12 manually constructed attacked cases derived from the controlled set. It targets known mechanisms without iterative target-model queries. Four triage cases add ticket or memory text designed to over-escalate finance-host compromise or portray administrator token abuse as benign maintenance. Four summarization cases introduce maintenance narratives and split instructions across ticket and retrieval evidence. Four response cases request monitoring-first treatment or delayed isolation through single-source, split, or persistent notes. These attacks preserve the original task schema and scorer.

The task-wise results are given in Table~\ref{tab:supp-adaptive}.

\begin{table}[H]
\centering
\begin{tabular}{lrrr}
\toprule
Task & $n$ & Raw TS & Matched TS\\
\midrule
Triage & 4 & 75.00\% & 100.00\%\\
Summary & 4 & 100.00\% & 100.00\%\\
Response & 4 & 100.00\% & 100.00\%\\
\bottomrule
\end{tabular}
\caption{Existing defense-aware results on \texttt{gpt-5.4}. Matched routes are severity protection for triage and firewalling for summary/response.}
\label{tab:supp-adaptive}
\end{table}

The undefended run fails one triage case, whereas the matched severity route solves all four; summary and response are unchanged. This slice therefore isolates the observed triage benefit and does not test IOC-specific adaptive attacks or query-budgeted optimization.

\end{document}